# MS2toImg: A Framework for Direct Bioactivity Prediction from Raw LC-MS/MS Data

Hansol Hong[1], Sangwon Lee[2,*], Jang-Ho Ha[2], Sung-June Chu[2], So-Hee An[2], Woo-Hyun Paek[2], Gyuhwa Chung[3], and Kyoung Tai No[1, 2*]

Affiliation
[1] Department of Integrative Biotechnology, Yonsei University, Incheon, 21983, Republic of Korea
[2] Bioinformatics & Molecular Design Research Center, Incheon, 21983, Republic of Korea
[3] Department of Biotechnology, Chonnam National University, Yeosu, 59626, Republic of Korea
[*] Correspondence Author
E-mail: swlee@bmdrc.org, ktno@yonsei.ac.kr

# Abstract

Untargeted metabolomics using LC-MS/MS offers the potential to comprehensively profile the chemical diversity of biological samples. However, the process is fundamentally limited by the "identification bottleneck," where only a small fraction of detected features can be annotated using existing spectral libraries, leaving the majority of data uncharacterized and unused. In addition, the inherently low reproducibility of LC-MS/MS instruments introduces alignment errors between runs, making feature alignment across large datasets both error-prone and challenging.

To overcome these constraints, we developed a deep learning method that eliminates the requirement for metabolite identification and reduces the influence of alignment inaccuracies. Here, we propose MS2toImg, a method that converts raw LC-MS/MS data into a two-dimensional images representing the global fragmentation pattern of each sample. These images are then used as direct input for a convolutional neural network (CNN), enabling end-to-end prediction of biological activity without explicit feature engineering or alignment.

Our approach was validated using wild soybean samples and multiple bioactivity assays (e.g., DPPH, elastase inhibition). The MS2toImg-CNN model outperformed conventional machine learning baselines (e.g., Random Forest, PCA), demonstrating robust classification accuracy across diverse tasks.

By transforming raw spectral data into images, our framework is inherently less sensitive to alignment errors caused by low instrument reproducibility, as it leverages the overall fragmentation landscape rather than relying on precise feature matching. This identification-free, image-based approach enables more robust and scalable bioactivity prediction from untargeted metabolomics data, offering a new paradigm for high-throughput functional screening in complex biological systems.

# Introduction

Natural products have provided humanity with a wealth of medicinal agents, chemical probes, and nutraceutical compounds, and continue to be a primary source of new drug leads and inspiration for synthetic chemistry[1], [2]. The study and discovery of bioactive small molecules from nature is an ever-evolving field, now steadily transformed by advances in analytical, computational, and biological science. With the broad adoption of untargeted metabolomics, particularly Liquid Chromatography-Mass Spectrometry (LC-MS), researchers are able to profile the chemical complexity of living systems at unprecedented depth and scale[3]. LC-MS is a key technique in metabolomics research, enabling the separation, quantification, and identification of metabolites with high selectivity and sensitivity[4]. Metabolomics research using LC-MS are categorized into targeted metabolomics, which quantifies a predefined set of metabolites, and untargeted metabolomics, which analyzes as many metabolites as possible without being limited to specific ones[5]. Untargeted metabolomics is widely studied for its ability to provide comprehensive insights into samples. This approach commonly employs tandem mass spectrometry (MS/MS), which provides structural information about metabolites and aids in distinguishing between them[6]. Untargeted metabolomics using LC-MS/MS facilitates comprehensive sample profiling by enabling both the quantitative and qualitative analysis of the global metabolome[7]. This approach does not rely on prior knowledge of the targeted chemicals, enabling rapid surveys of metabolic diversity across plants, microbes, marine organisms, and more.

However, despite these advantages, untargeted metabolomics faces fundamental methodological barriers that constrain its impact on natural products discovery and functional annotation. Chief among these barriers is the so-called "identification bottleneck," which refers to the inability to structurally identify or confidently annotate the vast majority of detected molecular features in complex extracts. In a typical untargeted LC-MS/MS experiment, thousands of precursor ions are detected, and only a fraction can be matched to known compounds using available reference spectral libraries[8]. As a result, most detected features remain unassigned, and much of the chemical and potential biological information in the sample is lost to downstream analysis.

Significant international efforts have been devoted to expanding reference resources for compound annotation, such as the Human Metabolome Database[9], Global Natural Products Social Molecular Networking[10], and advances in computational spectral prediction and in silico classification[11], [12], [13], [14]. Notably, tools like CANOPUS[15], MetDNA[16], and Spec2Vec[17] are allowing for impressive improvements in the computational annotation rate of MS/MS features. Nonetheless, these resources still cover only a small portion of the true chemical diversity found in complex extracts and are most successful at assigning compound classes, not unambiguous structural identities. In microbial and plant natural products research, the majority of bioactive molecules isolated each year are unknown to any public library, and bioassay-guided isolation workflows remain laborious and serial.

A second persistent challenge is related to instrumental reproducibility and feature alignment across samples and platforms[18]. LC-MS/MS data are affected by retention time drift[19], matrix effects[20], and mass calibration variability[21]. These variations, even when minimized by rigorous experimental protocols, can cause difficulties in matching features (i.e., the same metabolite across samples) and integrating peptide, lipid, or small molecule signals across broad datasets[22]. Consequently, feature tables constructed for downstream statistical analysis may be incomplete, redundant, or error-prone. Despite ongoing improvements in preprocessing software and batch effect correction, accurate feature alignment remains a nontrivial source of technical noise in metabolomics, especially as datasets scale up and diversity expands[23].

Traditional strategies for extracting biological meaning from untargeted datasets rely on multivariate analyses such as Principal Component Analysis (PCA), Principal Coordinates Analysis (PCoA)[24], and Orthogonal Partial Least Squares-Discriminant Analysis (OPLS-DA)[25]. These statistical tools are commonly used to visualize group separation and find features significantly associated with biological phenotypes. However, their utility is often hampered when the number of features greatly exceeds the number of samples, as is almost always the case in metabolomics. PCA may fail to resolve groups if most variance in the data is technical or unrelated to the biological comparison of interest. Supervised techniques like OPLS-DA appear to excel in class separation but often yielding overfitted models that perform well on training data but poorly on unseen samples[26]. These effects are exacerbated by the feature alignment challenges discussed above.

In response to these methodological obstacles, the field has increasingly looked toward deep learning and artificial intelligence for novel solutions. Recent studies have used deep learning to improve peak detection[27], [28], correct batch effects[29], predict retention time[30] and structure[11], [15], and automate other complex steps in metabolomics workflows. These advances suggest that data-driven models, freed from reliance on user-defined features, can both capture subtle non-linear associations and operate robustly in the presence of noise and technical variability. However, deep learning requires a large amount of high-quality data[31], and training on such data incurs high computational costs. There is thus a need for streamlined approaches that are both robust and computationally efficient.

To overcome the aforementioned challenges, studies have been conducted to represent mass spectrometry data as images and classify them using convolutional neural network (CNN) model, a type of deep learning model. Gonzalez et al. utilized CNN model to distinguish bacterial species based on lipid regions within the two-dimensional MS/MS data domain[32]. Similarly, Shen et al. transformed LC-MS raw data from human serum into images and applied a pre-trained CNN model to predict gestational age[33]. Although deep learning has been applied to mass spectrometry data in these studies, their scope remains limited. For instance, Gonzalez et al. utilized LC-MS/MS data but limited their analysis to lipid spectra, whereas Shen et al. examined broader profiles using LC-MS data, which, however, lack the rich fragmentation information available from MS/MS.

Advancing beyond these approaches, our study leverages the entirety of untargeted LC-MS/MS data for direct phenotype prediction in natural products research, presenting a novel, identification-free framework. Here, we introduce the MS2toImg framework, which encodes raw mass spectrometric fragmentation data from each sample into a single-channel, two-dimensional grayscale "molecular fingerprint" image. These images are then used as input for a CNN, eliminating the need for explicit feature selection, alignment, or annotation. Image-based approaches offer distinct advantages: they preserve the structural and spatial relationships between precursor and fragment ions, capture global data patterns, and are inherently robust to small translations and local noise due to the convolutional properties of deep networks[34].

In this study, we benchmarked PCA, OPLS-DA, Random Forest (RF), and a CNN using results from multiple functional bioassays on wild soybean (Glycine soja Sieb. & Zucc.), a valuable source of natural products[35]. We also optimized the input image resolution to balance predictive ability with computational efficiency. We propose a scalable and reproducible strategy for functional screening of complex natural product extracts by showing that highly accurate bioactivity prediction can be achieved from these grayscale molecular fingerprints alone, without needing prior chemical identification or extensive preprocessing.

## Results and Discussion

### Construction and Characteristics of MS/MS-Derived Molecular Fingerprint Images

We developed a workflow that transforms the raw MS/MS data into a grayscale image, forming an intuitive "molecular fingerprint". For each sample, MS-DIAL[36] was used to preprocess raw data and extract the precursor (MS1) and fragment (MS2 or MS/MS) m/z pairs along with their corresponding intensities. The intensity of the precursor ion was scaled between 0 and 1. The intensity value of the product ion was scaled between 0 and 1 with respect to the intensity of product ions belonging to the same precursor ion. Each value signifies the presence and relative abundance of a particular fragment derived from a specific precursor molecule in that sample. These values were then plotted onto a scatter plot, assigning the MS1 m/z to the x-axis and the MS2 m/z to the y-axis (Figure 1).

By representing the full spectrum of fragmentation events as spatially encoded pixel intensity values, this image-based strategy efficiently preserves the global structural information present in the original MS/MS data. Importantly, due to the inherent physics of tandem mass spectrometry, no fragment can have a greater m/z than its precursor. Therefore, the image region above the diagonal (MS2 > MS1) is structurally impossible and contains no meaningful data. To reduce the risk of overfitting and improve the interpretability of the model, we applied a masking procedure in which all pixels within this non-informative region were set to zero (black), ensuring that only data corresponding to possible fragmentation events are used for model training.

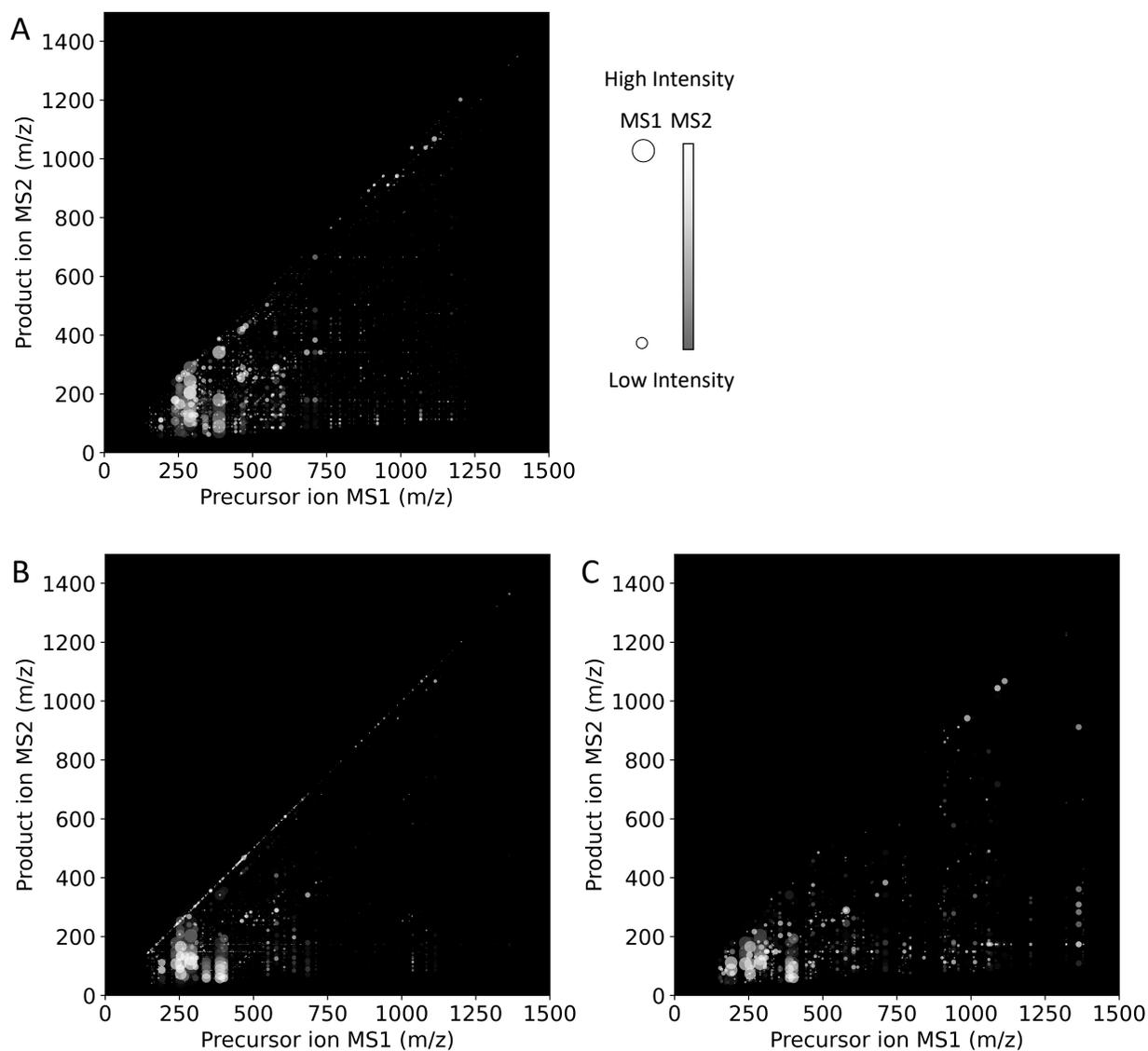

Figure 1. **Example images of MS/MS data converted to a scatter plot.** The plot correlates the mass-to-charge ratio (m/z) of precursor ions (MS1, x-axis) with their corresponding product ions (MS2 or MS/MS, y-axis). Each point represents a detected fragment ion from an MS/MS scan. The intensity of each ion is represented by both the point size and color saturation, providing a global overview of the fragmentation patterns present in the sample. The data is concentrated below the y=x diagonal, as fragments (MS2) necessarily have a smaller mass than their parent molecules (MS1). The figure illustrates how the complex tandem mass spectrometry data from three different samples—(A), (B), and (C)—are converted into information-rich images. Each image encapsulates the entirety of the sample's fragmentation behavior. The visual dissimilarities between the images highlight that each sample possesses a unique chemical signature, which can be learned by a convolutional neural network (CNN) to predict biological activity.

Pixel intensity in each grayscale image is proportional to the abundance of the corresponding fragment ion: low-intensity fragments appear as darker pixels, while fragments with high signal intensities appear brighter, nearing white. This visual representation serves a dual purpose: (i) it allows for direct, intuitive comparison of molecular fragmentation patterns across samples, and (ii) it supplies the convolutional neural network with high-dimensional yet spatially coherent inputs for data-driven feature extraction.

The advantages of this approach are multifold. First, it eliminates the need for explicit feature grouping, m/z alignment, or retention time correction, steps that are often error-prone and highly sensitive to technical variability across runs. Any minor run-to-run shifts in m/z values translate only into slight displacements of pixels within the image, which the CNN is robust to due to its convolutional operations and pattern recognition capacity. Second, the use of a grayscale image provides computational efficiency by minimizing data dimensionality while preserving the essential intensity-based information critical for classification. Lastly, this method enables the inclusion of all detected features—regardless of whether they can be annotated—directly into the predictive analysis, fully leveraging the information density of untargeted metabolomics.

An example molecular fingerprint image is presented in Figure 1, illustrating the high-density diagonal pattern reflecting abundant fragmentation and the clearly masked non-informative area. This data conversion process provides an end-to-end, identification-free pipeline, ensuring that none of the sample's chemical complexity is lost prior to machine learning-based bioactivity prediction.

## Model Architecture

To classify the grayscale "molecular fingerprint" images, we designed and optimized a Convolutional Neural Network (CNN), a class of deep learning models exceptionally well-suited for learning patterns from grid-like data such as images . The entire framework was implemented in Python (v3.10) using the Keras (v2.14) and TensorFlow libraries.

To ensure our model architecture was systematically optimized rather than arbitrarily chosen, we employed the Keras Tuner library to perform a random search across a wide hyperparameter space. This automated process explored various configurations, including the number of convolutional layers (1–3), the number of filters per layer (16–64), the number of dense layers (1–3), the number of units per dense layer (32–512), dropout rates (0.1–0.5), and the learning rate (logarithmically sampled from 1e-6 to 1e-4). The search was conducted for 120 trials, with the model configuration yielding the highest validation accuracy selected for final training and evaluation.

The final, optimized model architecture is depicted in Figure 2. It is composed of two main functional blocks:

1. A feature extraction block: This part consists of two sequential convolutional layers, each with a 3×3 kernel and ReLU activation function, followed by a 2×2 max-pooling layer. The convolutional layers act as learnable filters that scan the input image to detect hierarchical patterns, from simple edges (representing specific precursor-fragment relationships) to more complex motifs (representing broader chemical substructures). The max-pooling layers then downsample the feature maps, making the learned representations more robust to minor variations in the position of features within the image.

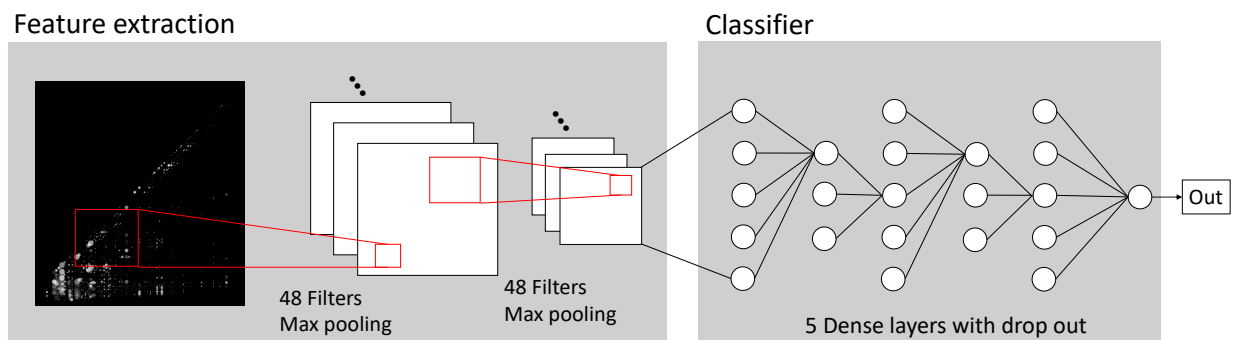

Figure 2. **The architecture of the Convolutional Neural Network (CNN) model.** The model consists of two main blocks: a feature extractor and a classifier. The feature extractor comprises two convolutional layers, each with 48 filters, followed by a max pooling layer for invariance to local translation. The subsequent classifier is composed of five fully connected dense layers with dropout regularization to prevent overfitting and map the learned features to the final output category.

2. A classification block: The output from the feature extractor is flattened into a one-dimensional vector and fed into a series of five fully connected dense layers. To prevent overfitting, a common challenge in models with many parameters, dropout regularization (rate = 0.5) was applied after each dense layer. The final output layer consists of a single neuron with a sigmoid activation function, which produces a probability score between 0 and 1 for binary classification.

The model was trained using the Adam optimizer and the binary cross-entropy loss function, standard choices for binary classification tasks. The selection of a CNN architecture is crucial, as its inherent properties directly address key challenges in metabolomics data. The convolutional operations provide a degree of translation invariance, meaning the model's ability to recognize a pattern is not strictly dependent on its exact location in the image. This directly translates to robustness against the minor instrumental drift (in m/z and retention time) that is unavoidable in LC-MS/MS experiments. Furthermore, by using single-channel grayscale images, we significantly reduce the computational complexity and number of trainable parameters compared to multi-channel inputs, leading to a more efficient and less overfitting-prone model without sacrificing essential predictive information.

## Impact of Image Resolution on Model Accuracy and Generalization

A critical hyperparameter in any image-based deep learning model is the resolution of the input data. This parameter directly influences the trade-off between the amount of information fed to the model and the risk of overfitting, as well as the overall computational cost. To identify the optimal input size for our "molecular fingerprint" images, we systematically evaluated the performance of the CNN model using three distinct resolutions: high (300×280 pixels), medium (150×140 pixels), and low (75×70 pixels).

The learning curves for each resolution, plotting training and validation accuracy over epochs, revealed a clear and informative trend (Table 1). The model trained on high-resolution images (Table 1 E, F) quickly achieved near-perfect accuracy on the training set. However, its performance on the validation set plateaued at a lower level, resulting in a significant gap between the two curves. This divergence is a classic indicator of overfitting, suggesting that at this high resolution, the model began to memorize noise and sample-specific artifacts rather than learning generalizable biological patterns.

Conversely, the model trained on low-resolution images (Table 1 A, B) exhibited underperformance. Both training and validation accuracies were significantly lower than those of the other models, indicating that the downsampling process had likely removed critical, discriminative features from the data, thereby limiting the model's predictive power.

Table 1. **MS2toImg-CNN model Performance Across Input Image Resolutions.** The training and validation curves for models trained on three different input image resolutions are shown: (A, B) low resolution, (C, D) medium resolution, and (E, F) high resolution. The left column (A, C, E) displays the loss, while the right column (B, D, F) displays the accuracy over 80 epochs. While higher resolution leads to lower training loss and near-perfect training accuracy (E, F), it also results in a larger gap between training and validation curves, indicating significant overfitting. The validation accuracy, which represents the model's true performance, peaks at the medium resolution (D) and does not improve further with high resolution (F). This suggests that medium resolution provides the optimal balance between feature preservation and model generalization, as excessively high resolution may cause the model to learn noise rather than meaningful patterns.

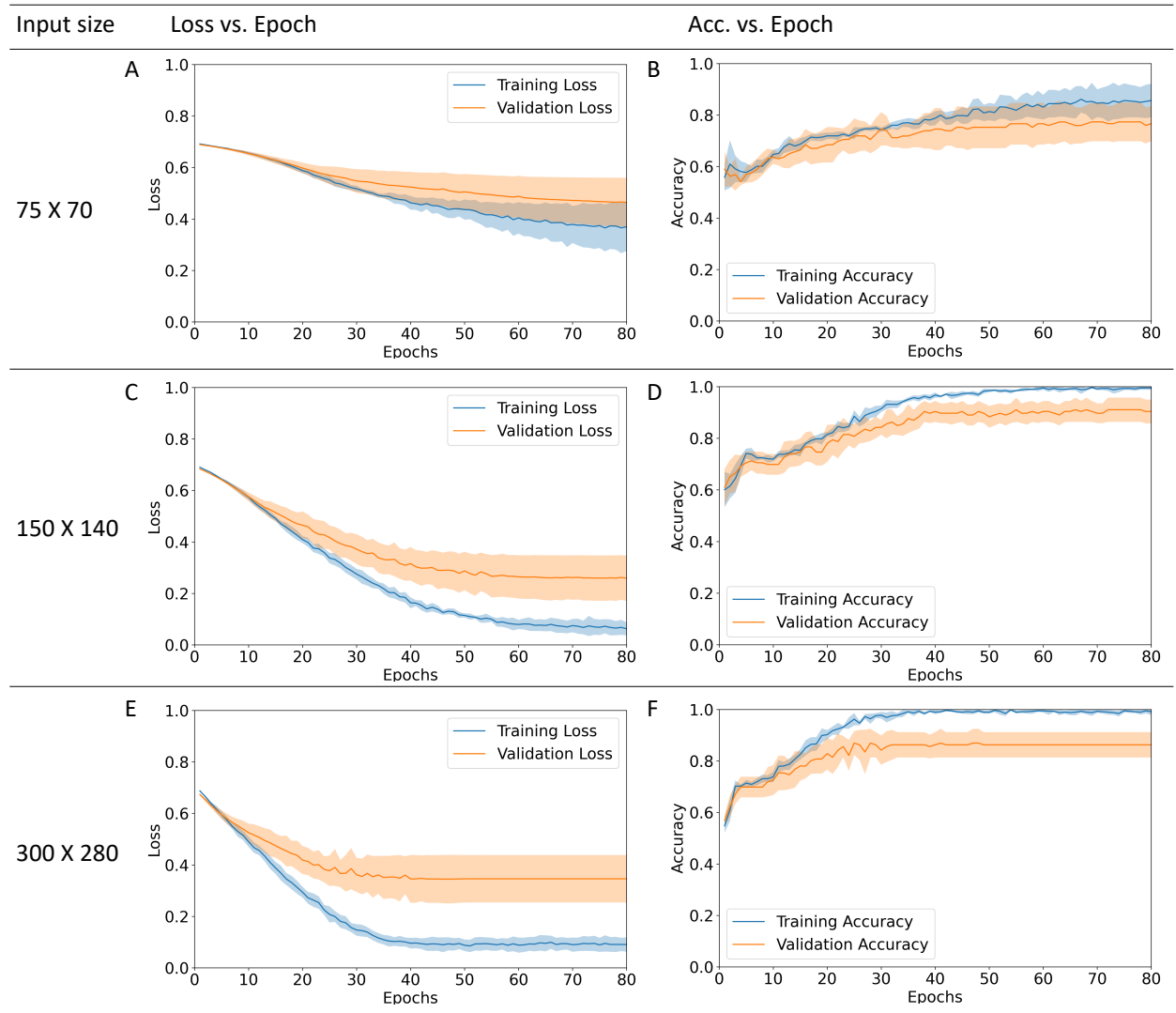

The medium-resolution images (Table 1 C, D) provided the optimal balance. This model achieved a high validation accuracy that was comparable to its training accuracy, indicating robust generalization without excessive overfitting. It successfully captured sufficient chemical information to make accurate predictions while being coarse-grained enough to avoid learning spurious noise.

This result underscores a crucial principle in machine learning: more data, in the form of higher resolution, is not always better. While higher resolution preserves more detail, it also increases model complexity and provides more opportunities for the model to learn irrelevant noise, ultimately harming its ability to generalize to unseen data. Our systematic evaluation allowed us to identify an optimal resolution that maximizes predictive performance while maintaining computational efficiency. Therefore, the 150×140 pixel resolution was selected and used for all subsequent experiments and comparisons described in this study.

## MS2toImg-CNN predicts the DPPH Antioxidant Activity

Having identified the optimal input resolution (150×140 pixels), we proceeded to evaluate the predictive performance of our CNN model using the DPPH radical scavenging assay as the primary bioactivity endpoint. DPPH is a widely used method for assessing antioxidant capacity and represents a biologically relevant phenotype that has been extensively studied in natural products research.

The optimized CNN model demonstrated excellent classification performance for distinguishing between high and low DPPH activity samples. A detailed analysis of individual sample predictions reveals strong concordance between the model's binary classifications and the experimental DPPH values (Figure 3A). The predicted labels (gray bars) show clear alignment with the measured antioxidant activity levels (red dots), with samples predicted as "High" consistently exhibiting elevated DPPH values and those classified as "Low" showing correspondingly reduced activity. This sample-by-sample comparison confirms that the model is learning meaningful biological patterns rather than random associations.

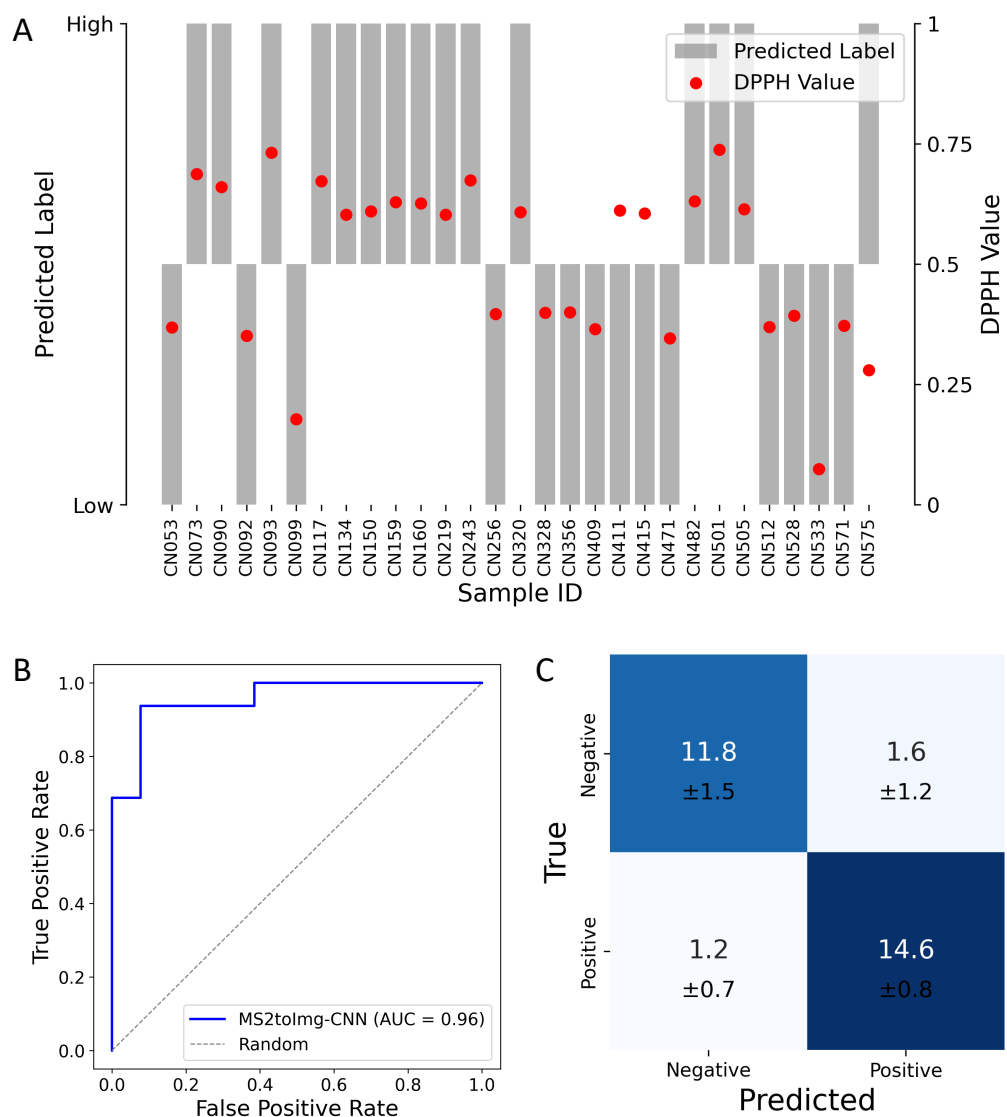

Figure 3. **Prediction performance of MS2toImg-CNN model. (A)** Predicted classification labels and experimental DPPH values for individual samples. The x-axis represents the sample IDs, the left y-axis shows the predicted binary label (High or Low antioxidant activity), and the right y-axis indicates the experimentally measured DPPH values. Gray bars denote model predictions, while red dots represent the corresponding DPPH assay values. This comparison illustrates the alignment between predicted classes and actual antioxidant activity levels. **(B)** Receiver operating characteristic (ROC) curve of MS2toImg-CNN model. The gray dashed line represents the performance baseline of a random classifier (AUC = 0.5). **(C)** Confusion matrix of model predictions obtained from 5-fold cross-validation. Values represent the mean number of samples per class (± standard deviation), averaged across folds. The heat map illustrates the distribution of true versus predicted labels, showing the model's ability to distinguish positive and negative classes.

To assess the model's discriminative capability, we analyzed the receiver operating characteristic (ROC) curve of our CNN model (Figure 3B). The model achieved an area under the curve (AUC) of 0.96, indicating superior discriminatory power across all decision thresholds. This high AUC value demonstrates that the CNN's end-to-end learning from molecular fingerprint images effectively captures relevant information for antioxidant activity prediction.

The robustness of the model's performance was further validated through 5-fold cross-validation analysis. The confusion matrix averaged across all folds (Figure 3C) shows balanced performance with minimal misclassification errors. The heat map visualization clearly illustrates the model's ability to distinguish between positive and negative classes, with the diagonal elements showing consistently high values and low standard deviations, indicating stable performance across different data splits. The low off-diagonal values confirm that false positives and false negatives are minimized, which is crucial for reliable screening applications in natural products discovery.

These comprehensive results establish that our image-based CNN framework can successfully learn complex, non-linear relationships between the global metabolic fingerprint captured in the grayscale images and the resulting antioxidant activity. The model's ability to achieve such high performance without any prior knowledge of chemical identities or biomarker compounds demonstrates the power of this identification-free approach. This strong performance on DPPH prediction provided the foundation for evaluating the framework's generalizability across additional bioassays, as described in subsequent sections.

# Comparative Performance of MS2toImg-CNN and Conventional Methods

To evaluate the impact of our image-based data representation strategy, we compared the performance of the MS2toImg-CNN framework with several conventional analytical methods. The MS2toImg-CNN model was trained on two-dimensional grayscale "molecular fingerprint" images, whereas PCA, OPLS-DA, and RF were trained on the traditional spectral data (alignment-dependent feature table).

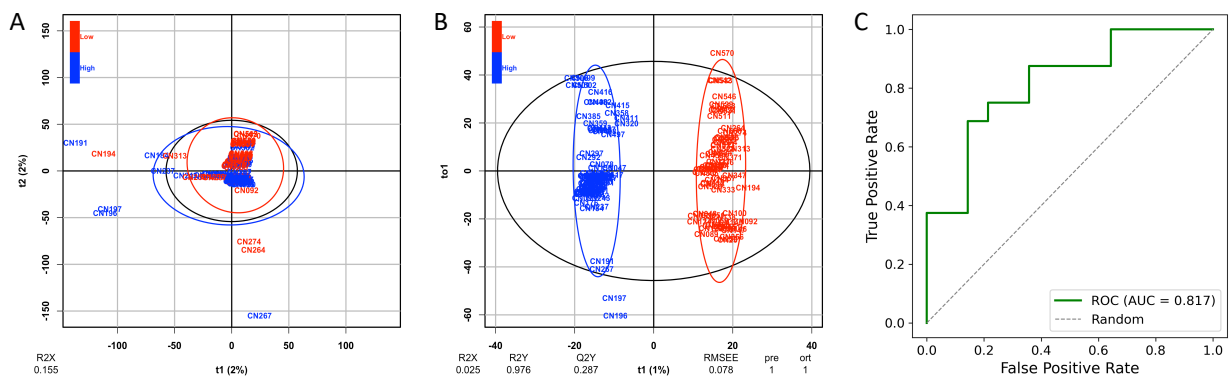

Figure 4. **Evaluation of PCA, OPLS-DA, and RF Models.** All three models used spectral data as input. (A) In the unsupervised PCA scores plot, no clear separation was observed between the "High" (red) and "Low" (blue) groups, with substantial overlap indicating that intrinsic data variance alone is insufficient to discriminate between the two classes. (B) In contrast, the supervised OPLS-DA scores plot achieved visually perfect separation, as also reflected by a high $R^2Y$ value (0.976). However, the model's predictive ability, evaluated by cross-validation, was extremely poor ($Q^2Y = 0.287$), clearly demonstrating that the OPLS-DA model was severely overfitted and lacked practical predictive power. (C) The ROC curve of the RF classifier yielded an AUC of 0.817, indicating a reasonably generalized classification performance.

First, we applied unsupervised PCA to the feature table. The resulting scores plot showed no clear separation between the "High" and "Low" bioactivity groups, indicating that the principal sources of variance in the feature data were not correlated with the biological phenotype of interest (Figure 4A). In contrast, the supervised method OPLS-DA yielded a model with visually perfect separation and a high goodness-of-fit value ($R^2Y > 0.97$). However, its predictive ability, assessed by seven-fold cross-validation, was extremely poor ($Q^2Y < 0.28$) (Figure 4B). This pronounced discrepancy between $R^2Y$ and $Q^2Y$ is a classic indicator of severe overfitting when applied to high-dimensional metabolomics feature tables, rendering such models unreliable for this classification task.

Next, we compared our MS2toImg-CNN model with an RF classifier trained on the same feature table. As shown by the ROC curves, the MS2toImg-CNN model (AUC = 0.96; Figure 3B) exhibited markedly superior discriminative performance relative to the RF model (AUC = 0.817; Figure 4C). Moreover, as illustrated in Figure 5, the MS2toImg-CNN model achieved superior or comparable results across all four standard evaluation metrics: accuracy, recall, F1 score, and precision. The most significant advantage of our MS2toImg-CNN approach was observed in recall (Figure 5B), where the model demonstrated a substantially higher ability to correctly identify true positive samples. This is particularly important in screening applications, where missing a positive case (false negative) is often more detrimental than generating a false positive. Consequently, the high recall contributed to an improved F1 score (Figure 5C), which reflects a balanced integration of precision and recall.

The superior performance of the MS2toImg-CNN model can be attributed to the inherent robustness of the image-based framework against instrumental variation. Traditional methods such as RF are highly dependent on the quality of the pre-processed feature table, and any errors or inconsistencies introduced during feature detection, grouping, or alignment propagate through the analysis and degrade performance. In contrast, our MS2toImg-CNN directly learns discriminative features from the global data landscape of the "molecular fingerprint" image. Minor run-to-run variations in retention time or m/z, which often confound alignment algorithms, simply result in small pixel shifts within the image. The convolutional filters of the CNN are inherently tolerant of such spatial variations, enabling the model to reliably recognize key patterns despite instrumental noise. Collectively, these results demonstrate that our framework is not only more accurate but also fundamentally more robust and better suited for analyzing real-world, variable metabolomics data.

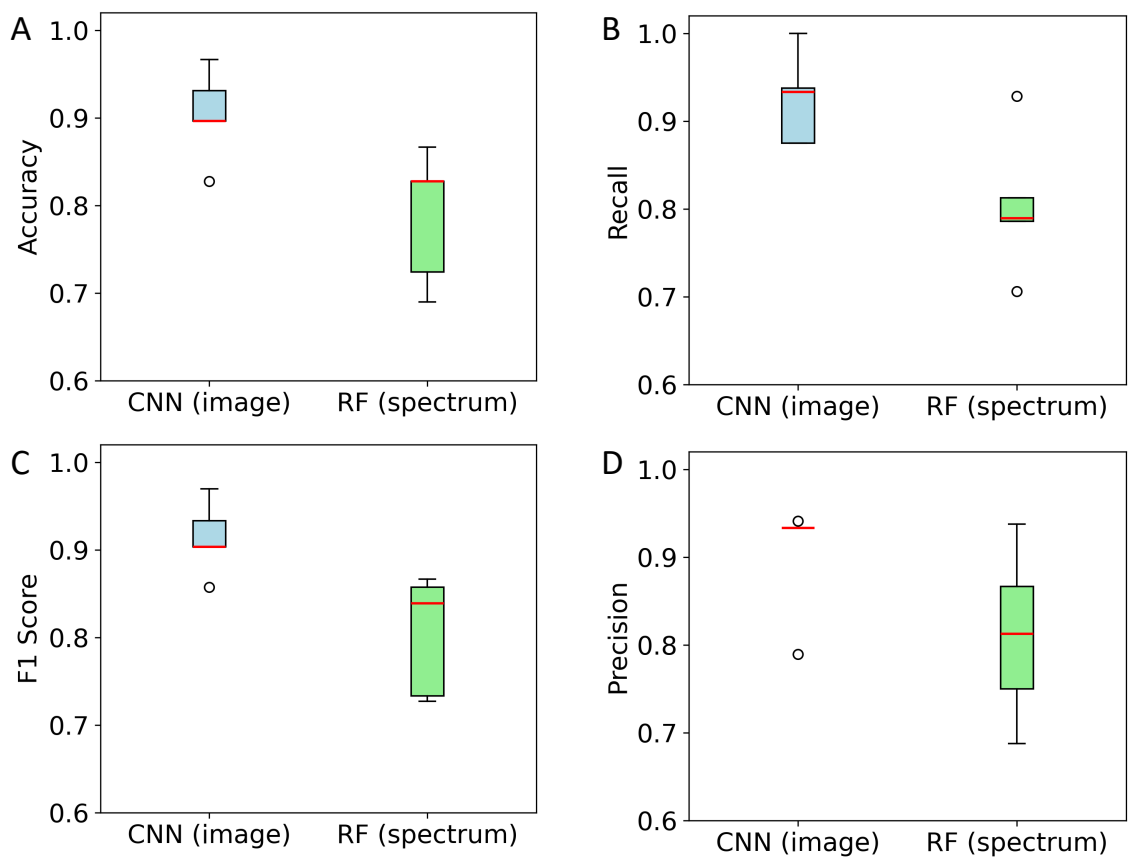

Figure 5. **Prediction Performance metrics of CNN and RF.** The performance of the two models was compared using standard evaluation metrics. The CNN model was trained directly on image data, whereas the RF model was trained on spectral data. The box plots represent the average scores for (A) Accuracy, (B) Recall, (C) Precision, and (D) F1 Score. Overall, the CNN exhibited higher recall, resulting in fewer missed positive cases, and consequently achieved superior or slightly higher F1 scores compared to the RF model.

## Generalizability of the MS2toImg-CNN model

A key measure of a novel analytical framework's utility is its ability to generalize across different tasks and datasets without extensive re-engineering. To assess the versatility and broader applicability of our MS2toImg-CNN approach, we applied the identical, optimized framework to predict three additional, distinct biological activities measured on the same sample set: elastase inhibition (an anti-wrinkle indicator), total flavonoid content (TFC), and total phenolic content (TPC). The same data conversion process and CNN architecture were used for each task, with the only change being the biological endpoint labels used for training.

The MS2toImg framework demonstrated robust and consistent predictive performance across all tested bioassays. As summarized in Table 2, the model achieved high classification accuracies on the independent test sets for elastase inhibition, TFC, and TPC, comparable to the performance observed for the initial DPPH assay. This consistent success is particularly noteworthy because each assay probes a different biological mechanism, which is likely driven by different sets of underlying bioactive compounds within the complex natural product extracts.

This result strongly indicates that our model is not simply memorizing a few specific biomarkers for a single task. Instead, it is learning a more fundamental and transferable representation of the relationship between the global metabolic fingerprint and its resulting biological function. This fingerprint is captured within the 2D image. The ability to successfully predict outcomes for mechanistically diverse assays confirms that the image-based features are rich and informative enough to capture a wide range of structure-activity relationships.

Table 2. **Summary of model performance across different assay types.** The table presents average classification accuracy and standard deviation (SD) for assays including Elastase Inhibitory Activity (anti-wrinkle), Total Flavonoid Content (TFC), and Total Phenolic Content (TPC).

| Assay type | Loss vs. Epoch | Acc. vs. Epoch | Test set (avg.) |
|---|---|---|---|
| Elastase Inhibitory Activity (Anti-wrinkle) | 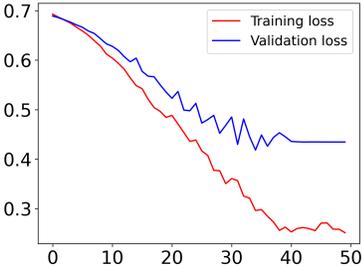 | 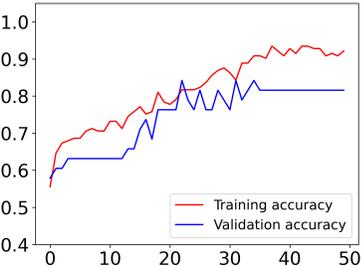 | Accuracy: 0.79<br>SD: 0.055 |
| TFC | 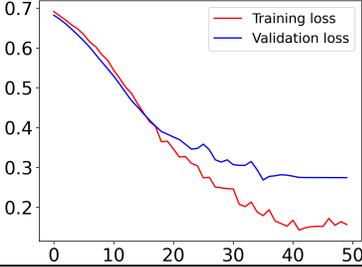 | 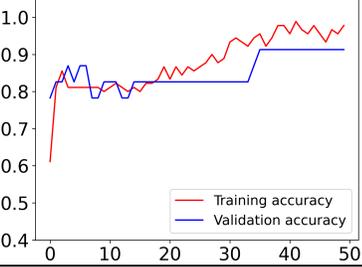 | Accuracy: 0.88<br>SD: 0.11 |
| TPC | 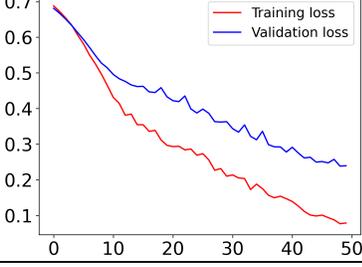 | 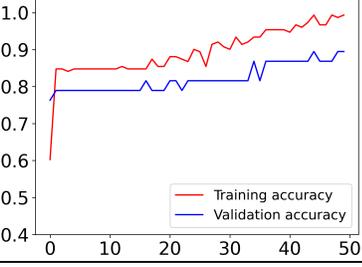 | Accuracy: 0.88<br>SD: 0.031 |

# Conclusion

In this study, we successfully developed and validated a novel deep learning framework that addresses two fundamental limitations in untargeted metabolomics: the identification bottleneck and the challenges of instrumental reproducibility. By converting raw LC-MS/MS data into two-dimensional grayscale "molecular fingerprint" images and using them as direct input for a Convolutional Neural Network, we have established a new paradigm for identification-free bioactivity prediction. Our approach effectively leverages the entire chemical profile of a sample, including both known and unknown compounds, and its image-based nature provides inherent robustness against the minor run-to-run variations that confound traditional alignment-based methods.

The remarkable performance of our CNN model compared to conventional machine learning approaches, coupled with its proven generalizability across multiple distinct bioassays, highlights its potential as a powerful, computationally efficient, and scalable tool. This framework opens new possibilities for high-throughput functional screening in complex biological systems, shifting the focus of metabolomics from exhaustive chemical annotation towards direct, data-driven phenotypic interpretation.

# Methods

## Sample preparation

Wild soybeans (*Glycine soja* Sieb. & Zucc.) were washed with distilled water and dried at room temperature. The dried wild soybeans were ground using a disposable grinding chamber (MT 40, IKA, Staufen, Germany) at 20,000 rpm for 1 minute, repeated three times. The ground samples were extracted twice at room temperature for 15 minutes each using 70% methanol at a volume 20 times that of the sample in an ultrasonic bath (Power Sonic 420; Hwashin Tech Co., Ltd., Seoul, Korea). The resulting extracts were centrifuged at 3,500 rpm for 5 minutes (CRYSTE VARISPIN4; NOVAPRO Co., Ltd., Gwangmyeong, Republic of Korea), and the supernatants were collected.

The collected supernatants were filtered through Whatman no. 1 filter paper (Whatman International Ltd., Maidstone, UK) using a Büchner funnel, and the filtrates were concentrated using a rotary evaporator (NVP-2100V, EYELA, Tokyo, Japan) in a 50°C water bath. To remove proteins, cold acetone was added at four times the volume of the concentrated extract, thoroughly mixed, and stored at –20°C for over 12 hours. The mixture was then centrifuged at 3,000 rpm for 5 minutes, and the supernatant was collected and evaporated using a rotary evaporator, followed by lyophilization.

For lipid removal, the lyophilized powder was dissolved in 80% methanol (2 mg/mL), and hexane at twice the volume of the extract solution was added. The 80% methanol layer was separated and collected after three repetitions of the extraction process. The collected fraction was concentrated using a rotary evaporator. The final lipid-removed extract was dissolved in MS-grade methanol at a concentration of 1 mg/mL, filtered through a 0.2 μm PTFE syringe filter (DISMIC®; Advantec MFS, Inc., Dublin, CA, USA), and transferred into vials for UHPLC-MS/MS analysis. The prepared extracts were also used for the DPPH radical scavenging assay.

## DPPH Radical Scavenging Assay

The extract was prepared at a concentration of 0.5 mg/mL. A standard calibration curve was constructed using ascorbic acid at concentrations of 20, 40, 60, 80, 100, 120, 140, and 160 μg/mL. The prepared extract and standard solutions were mixed with DPPH solution (0.3 mM 2,2-diphenyl-1-picrylhydrazyl in 80% methanol; optical density: 0.7 ± 0.1) at a 1:19 (v/v) ratio and dispensed into a 96-well plate. The mixtures were incubated at room temperature in the dark for 30 minutes. After incubating, absorbance was measured at 517 nm. All measurements were performed in triplicate. The DPPH radical scavenging activity (%) was calculated using the following equation:

$$\text{Radical scavenging activity}(\%) = \frac{\text{Blank O.D} - \text{Sample O.D}}{\text{Blank O.D}} \times 100$$

where Blank OD is the absorbance of the control after the reaction, and Sample OD is the absorbance of the sample after the reaction.

## LC-MS/MS Analysis

The prepared extracts (1 mg/mL) were separated using ultra-high-performance liquid chromatography (UHPLC, Vanquish Horizon, Thermo Fisher Scientific) and detected with an Orbitrap ID-X Tribrid mass spectrometer (Thermo Fisher Scientific). The analytical column used was Hypersil GOLD™ Vanquish (150 × 2.1 mm, 1.9 μm; Thermo Fisher Scientific), and the analysis was performed in triplicate. The mobile phase consisted of water as solvent A and acetonitrile containing 0.1% formic acid as solvent B. The UHPLC conditions were as follows: 0–3 min (5% B), 3–7 min (5–11.5% B), 7–8 min (11.5–12% B), 8–15 min (12–14% B), 15–25 min (14–25% B), 25–27 min (25–50% B), 27–29 min (50% B), 29–34 min (50–60% B), 34–36 min (60–100% B), and 36–50 min (100% B). The flow rate was set at 0.3 mL/min. Samples of 1 μL were injected into the column, and the column temperature was maintained at 45 °C. The ESI source conditions were set as follows: negative ion spray voltage at 2500 V, sheath gas flow rate at 50 Arb, auxiliary gas flow rate at 10 Arb, sweep gas flow rate at 1 Arb, ion transfer tube temperature at 320 °C, and vaporizer temperature at 300 °C. The full MS scan resolution was set at 60,000 with a scan range of 140–2000 Da and an RF lens was set to 35%. The stepped HCD collision energies for the ddMS/MS scan were set at 30%, 40%, and 50%, with a resolution of 15,000.

## Data Preprocessing

Raw mass spectrometry (MS) data files were preprocessed using MS-DIAL software. Upon launching MS-DIAL, the data type for MS1 was set to "profile," while the data type for MS/MS was set to "centroid." Triplicate measurements of each sample were aligned using blank and quality control (QC) samples, with the QC sample designated as the reference. Retention time tolerance for alignment was set to 1 minute. The following adduct ions were selected in the adduct settings: $[M-H]^-$, $[M-H_2O-H]^-$, $[M+Na-2H]^-$, $[M-K-2H]^-$, and $[M-FA-H]^-$. Following peak alignment, peak intensities were normalized using the total ion chromatogram (TIC) method.

Principal component analysis (PCA) was then conducted via the data visualization module to identify potential outliers. After verifying outliers and confirming normalization, the aligned data were exported as a text file using the MS-DIAL export function. Only peak intensity values greater than $10^{-4}$ were retained for further analysis.

## Conversion of LC-MS/MS data into 2D Images

The images used as input to the CNN model were generated in the following ways. A total of four information were extracted from the txt file, the result file of MS-DIAL: 1) the m/z value of precursor ion (MS1) 2) the intensity value of precursor ion 3) the m/z values of product ions (MS2), and 4) the intensity values of product ions. The intensity values of the precursor ions were normalized to a range between 0 and 1. The intensity values of the product ions were scaled between 0 and 1 with respect to the intensity of the product ions belonging to the same precursor ion.

Each image was generated as a scatter plot using matplotlib 3.8.4 in python 3.12.3. In the plot, MS1 m/z values were placed on the x-axis (100 ~ 1500 m/z), and the precursor ion intensities were represented by the size of each point. MS2 m/z values were plotted on the y-axis (0 ~ 1500 m/z), and product ion intensities were encoded as grayscale color intensities. All images were rendered at 600 dpi. Images of sizes 75 × 70, 150 × 140, and 300 × 280, used for model training, were generated using the image resize function in TensorFlow 2.15.

## CNN Architecture and Training Procedure

A CNN model was employed to classify antioxidant activity based on features extracted from spectral images. The model was implemented in Python 3.10 using Tensorflow 2.15.

The network architecture consisted of two convolutional layers followed by a max pooling layer for feature extraction. Each convolutional layer included 48 filters with a kernel size of 3 × 3. Feature representations from the convolutional layers were passed through five fully connected (dense) layers, each followed by a dropout layer to prevent overfitting. The ReLU activation function was applied to all hidden layers, and the Adam optimizer was used for model optimization. The output layer employed a sigmoid activation function for binary classification, with binary cross-entropy used as the loss function. The hyperparameters used for training are summarized in Supplementary Table 1.

All DPPH radical scavenging values were normalized to a range between 0 and 1. Samples with normalized values below 0.4 were categorized as the low-activity group, while those above 0.6 were categorized as the high-activity group. Both high- and low-activity samples were used to train the model. To reduce model bias and maximize data utilization, fivefold cross-validation was performed using the StratifiedKFold function from scikit-learn version 1.3.0.

# Feature Table Construction for Multivariate analysis and Random Forest

To enable Random Forest (RF) classification and multivariate analysis, all detected features across the 545 wild soybean accessions were merged into a unified feature table. The construction of this table involved two sequential merging steps: first, the consolidation of redundant features within individual samples; and second, the merging of common features across different samples. The merging criteria were based on similarity scores calculated for mass-to-charge ratio (MS1), retention time (RT), and MS/MS spectral similarity.

The similarity scores for MS1 and RT were calculated using the following equation[37]:

$$\text{MS1 or RT similarity} = \exp\left(-0.5 \cdot \left(\frac{\text{experimental value}_1 - \text{experimental value}_2}{\sigma}\right)^2\right)$$

where the sigma ($\sigma$) parameter was defined as follows: $\sigma$ for RT was set to the search tolerance value, while $\sigma$ for MS1 was set to the empirical standard deviation of the observed MS1 values. In this equation, the similarity score decreases rapidly as the difference between the two values increases. A score close to 1 indicates strong similarity, whereas a score near 0 indicates low or no similarity.

MS/MS spectral similarity was evaluated using two different methods: cosine similarity and spectral entropy. Cosine similarity quantifies the angular distance between MS/MS spectral vectors, focusing on the alignment of fragment ion patterns[38]. In contrast, MS/MS spectral similarity was evaluated using spectral entropy, an information-theoretic approach that treats the MS/MS spectrum as a probability distribution[39]. This method accounts for both MS1 and intensity distributions and quantifies the divergence between spectra based on their entropy differences. When the average of the three similarity scores (MS1, RT, MS/MS) exceeded 0.8, the features were considered equivalent and merged; otherwise, they were treated as distinct and added as new entries to the feature table. All merging procedures were implemented in Python 3.12.3 using pandas 2.2.2.

As a result, one feature table was generated using cosine similarity and the other using spectral entropy. Both feature tables were subsequently used to evaluate the effect of MS/MS similarity metrics on downstream classification performance.

## Conventional Multivariate Methods and Random Forest

To evaluate the classification performance of conventional machine learning and multivariate statistical approaches, we applied Random Forest (RF), principal component analysis (PCA), and orthogonal partial least squares discriminant analysis (OPLS-DA). RF classification was performed using two types of feature tables generated by different MS/MS similarity metrics: cosine similarity and spectral entropy. The RF models were implemented using scikit-learn 1.3.0, with default hyperparameters unless otherwise stated. Model performance was assessed based on accuracy, precision, recall, and F1-score using 5-fold cross-validation.

For unsupervised analysis, PCA was conducted to explore overall variance in the dataset. Samples with high and low DPPH scavenging activity were projected onto the first two principal components to assess natural grouping tendencies. In addition, OPLS-DA was performed as a supervised method to visualize class separation. The OPLS-DA model was evaluated using $R^2Y$ and $Q^2Y$ values, and model validity was further assessed using permutation tests.

These conventional approaches provided baseline comparisons to the CNN-based image classification model. While PCA and OPLS-DA offered limited predictive utility, RF model served as a robust non-deep learning classifier. However, all conventional methods required prior feature merging process, in contrast to the image-based CNN model which eliminated the need for manual preprocessing.

PCA and OPLS-DA were conducted using the ropls package (v1.32.0) in RStudio (v2023.06.1+524).

# Data availability

The raw data generated during this study are available from the corresponding author on reasonable request.

# Code availability

The MS2toImg-CNN source code and data used to generate the results and figures in this paper are available on GitHub repository at: https://github.com/jsehhs/MS2toImg-CNN.

# Acknowledgements

This research was supported by the Ministry of Trade, Industry and Energy (MOTIE), Korea, under the "Infrastructure Support Program for Industry Innovation" (Reference No. P0014714).

# Author Contributions

H. Hong: Conceptualization, Formal Analysis, Software, Writing – Original Draft, Writing – Review & Editing, Visualization. S. Lee: Conceptualization, Software, Validation, Writing – Review & Editing. J.-H. Ha: Investigation. S.-J. Chu: Investigation. S.-H. An: Investigation. W.-H. Paek: Investigation. G. Chung: Resources. K. T. No: Supervision, Project Administration, Funding Acquisition.

# References

[1]  D. J. Newman and G. M. Cragg, 'Natural products as sources of new drugs over the 30 years from 1981 to 2010', Mar. 23, 2012. doi: 10.1021/np200906s.

[2]  S. Xie, F. Zhan, J. Zhu, S. Xu, and J. Xu, 'The latest advances with natural products in drug discovery and opportunities for the future: a 2025 update', *Expert Opin Drug Discov*, 2025, doi: 10.1080/17460441.2025.2507382.

[3]  J. L. Ren, A. H. Zhang, L. Kong, and X. J. Wang, 'Advances in mass spectrometry-based metabolomics for investigation of metabolites', *RSC Adv*, vol. 8, no. 40, pp. 22335–22350, 2018, doi: 10.1039/c8ra01574k.

[4]  J. H. Gross, 'Tandem Mass Spectrometry', in *Mass Spectrometry: A Textbook*, Cham: Springer International Publishing, 2017, pp. 539–612. doi: 10.1007/978-3-319-54398-7_9.

[5]  H. Gika, C. Virgiliou, G. Theodoridis, R. S. Plumb, and I. D. Wilson, 'Untargeted LC/MS-based metabolic phenotyping (metabonomics/metabolomics): The state of the art', Jun. 01, 2019, *Elsevier B.V.* doi: 10.1016/j.jchromb.2019.04.009.

[6]  M. T. Sheldon, R. Mistrik, and T. R. Croley, 'Determination of Ion Structures in Structurally Related Compounds Using Precursor Ion Fingerprinting', *J Am Soc Mass Spectrom*, vol. 20, no. 3, pp. 370–376, Mar. 2009, doi: 10.1016/j.jasms.2008.10.017.

[7]  G. Theodoridis, H. G. Gika, and I. D. Wilson, 'LC-MS-based methodology for global metabolite profiling in metabonomics/metabolomics', *TrAC - Trends in Analytical Chemistry*, vol. 27, no. 3, pp. 251–260, Mar. 2008, doi: 10.1016/j.trac.2008.01.008.

[8]  R. R. Da Silva, P. C. Dorrestein, and R. A. Quinn, 'Illuminating the dark matter in metabolomics', Oct. 13, 2015, *National Academy of Sciences*. doi: 10.1073/pnas.1516878112.

[9] D. S. Wishart *et al.*, 'HMDB 5.0: The Human Metabolome Database for 2022', *Nucleic Acids Res*, vol. 50, no. D1, pp. D622–D631, Jan. 2022, doi: 10.1093/nar/gkab1062.

[10] M. Wang *et al.*, 'Sharing and community curation of mass spectrometry data with Global Natural Products Social Molecular Networking', Sep. 08, 2016, *Nature Publishing Group*. doi: 10.1038/nbt.3597.

[11] S. M. Colby, J. R. Nuñez, N. O. Hodas, C. D. Corley, and R. R. Renslow, 'Deep Learning to Generate in Silico Chemical Property Libraries and Candidate Molecules for Small Molecule Identification in Complex Samples', *Anal Chem*, vol. 92, no. 2, pp. 1720–1729, Jan. 2020, doi: 10.1021/acs.analchem.9b02348.

[12] F. Huber, S. van der Burg, J. J. J. van der Hooft, and L. Ridder, 'MS2DeepScore: a novel deep learning similarity measure to compare tandem mass spectra', *J Cheminform*, vol. 13, no. 1, Dec. 2021, doi: 10.1186/s13321-021-00558-4.

[13] H. Ji, Y. Xu, H. Lu, and Z. Zhang, 'Deep MS/MS-Aided Structural-Similarity Scoring for Unknown Metabolite Identification', *Anal Chem*, vol. 91, no. 9, pp. 5629–5637, May 2019, doi: 10.1021/acs.analchem.8b05405.

[14] Y. Li, T. Kind, J. Folz, A. Vaniya, S. S. Mehta, and O. Fiehn, 'Spectral entropy outperforms MS/MS dot product similarity for small-molecule compound identification', *Nat Methods*, vol. 18, no. 12, pp. 1524–1531, Dec. 2021, doi: 10.1038/s41592-021-01331-z.

[15] K. Dührkop *et al.*, 'Systematic classification of unknown metabolites using high-resolution fragmentation mass spectra', *Nat Biotechnol*, vol. 39, no. 4, pp. 462–471, Apr. 2021, doi: 10.1038/s41587-020-0740-8.

[16]  Z. Zhou, M. Luo, H. Zhang, Y. Yin, Y. Cai, and Z. J. Zhu, 'Metabolite annotation from knowns to unknowns through knowledge-guided multi-layer metabolic networking', *Nat Commun*, vol. 13, no. 1, Dec. 2022, doi: 10.1038/s41467-022-34537-6.

[17]  F. Huber *et al.*, 'Spec2Vec: Improved mass spectral similarity scoring through learning of structural relationships', *PLoS Comput Biol*, vol. 17, no. 2, Feb. 2021, doi: 10.1371/JOURNAL.PCBI.1008724.

[18]  A. M. Năstase *et al.*, 'Alignment of multiple metabolomics LC-MS datasets from disparate diseases to reveal fever-associated metabolites', *PLoS Negl Trop Dis*, vol. 17, no. 7 July, Jul. 2023, doi: 10.1371/journal.pntd.0011133.

[19]  C. T. Wu *et al.*, 'Targeted realignment of LC-MS profiles by neighbor-wise compound-specific graphical time warping with misalignment detection', *Bioinformatics*, vol. 36, no. 9, pp. 2862–2871, May 2020, doi: 10.1093/bioinformatics/btaa037.

[20]  A. K. Boysen, K. R. Heal, L. T. Carlson, and A. E. Ingalls, 'Best-Matched Internal Standard Normalization in Liquid Chromatography-Mass Spectrometry Metabolomics Applied to Environmental Samples', *Anal Chem*, vol. 90, no. 2, pp. 1363–1369, Jan. 2018, doi: 10.1021/acs.analchem.7b04400.

[21]  F. Jiang *et al.*, 'Signal Drift in Liquid Chromatography Tandem Mass Spectrometry and Its Internal Standard Calibration Strategy for Quantitative Analysis', *Anal Chem*, vol. 92, no. 11, pp. 7690–7698, Jun. 2020, doi: 10.1021/acs.analchem.0c00633.

[22]  J. D. Watrous *et al.*, 'Visualization, Quantification, and Alignment of Spectral Drift in Population Scale Untargeted Metabolomics Data', *Anal Chem*, vol. 89, no. 3, pp. 1399–1404, Feb. 2017, doi: 10.1021/acs.analchem.6b04337.

[23] J. Guo and T. Huan, 'Mechanistic Understanding of the Discrepancies between Common Peak Picking Algorithms in Liquid Chromatography-Mass Spectrometry-Based Metabolomics', *Anal Chem*, vol. 95, no. 14, pp. 5894–5902, Apr. 2023, doi: 10.1021/acs.analchem.2c04887.

[24] F. Di Ottavio *et al.*, 'A UHPLC-HRMS based metabolomics and chemoinformatics approach to chemically distinguish "super foods" from a variety of plant-based foods', *Food Chem*, vol. 313, May 2020, doi: 10.1016/j.foodchem.2019.126071.

[25] A. Mie *et al.*, 'Discrimination of conventional and organic white cabbage from a long-term field trial study using untargeted LC-MS-based metabolomics', *Anal Bioanal Chem*, vol. 406, no. 12, pp. 2885–2897, 2014, doi: 10.1007/s00216-014-7704-0.

[26] R. Bujak, E. Daghir-Wojtkowiak, R. Kaliszan, and M. J. Markuszewski, 'PLS-based and regularization-based methods for the selection of relevant variables in non-targeted metabolomics data', *Front Mol Biosci*, vol. 3, no. JUL, Jul. 2016, doi: 10.3389/fmolb.2016.00035.

[27] A. D. Melnikov, Y. P. Tsentalovich, and V. V. Yanshole, 'Deep Learning for the Precise Peak Detection in High-Resolution LC-MS Data', *Anal Chem*, vol. 92, no. 1, pp. 588–592, Jan. 2020, doi: 10.1021/acs.analchem.9b04811.

[28] Y. Gloaguen, J. A. Kirwan, and D. Beule, 'Deep Learning-Assisted Peak Curation for Large-Scale LC-MS Metabolomics', *Anal Chem*, vol. 94, no. 12, pp. 4930–4937, Mar. 2022, doi: 10.1021/acs.analchem.1c02220.

[29] Z. Rong *et al.*, 'NormAE: Deep Adversarial Learning Model to Remove Batch Effects in Liquid Chromatography Mass Spectrometry-Based Metabolomics Data', *Anal Chem*, vol. 92, no. 7, pp. 5082–5090, Apr. 2020, doi: 10.1021/acs.analchem.9b05460.

[30] Q. Yang, H. Ji, H. Lu, and Z. Zhang, 'Prediction of Liquid Chromatographic Retention Time with Graph Neural Networks to Assist in Small Molecule Identification', *Anal Chem*, vol. 93, no. 4, pp. 2200–2206, Feb. 2021, doi: 10.1021/acs.analchem.0c04071.

[31] 'Why the metabolism field risks missing out on the AI revolution', *Nat Metab*, vol. 1, no. 10, pp. 929–930, 2019, doi: 10.1038/s42255-019-0133-9.

[32] L. E. Gonzalez *et al.*, 'Machine-Learning Classification of Bacteria Using Two-Dimensional Tandem Mass Spectrometry', *Anal Chem*, vol. 95, no. 46, pp. 17082–17088, Nov. 2023, doi: 10.1021/acs.analchem.3c04016.

[33] B. Sun *et al.*, 'The combination of deep learning and pseudo-MS image improves the applicability of metabolomics to congenital heart defect prenatal screening', *Talanta*, vol. 275, Aug. 2024, doi: 10.1016/j.talanta.2024.126109.

[34] A. Bietti and J. Mairal, 'Group Invariance, Stability to Deformations, and Complexity of Deep Convolutional Representations', 2019. [Online]. Available: http://jmlr.org/papers/v20/18-190.html.

[35] M. A. Nawaz *et al.*, 'Korean wild soybeans (glycine soja sieb and zucc.): Geographic distribution and germplasm conservation', Feb. 02, 2020, *MDPI*. doi: 10.3390/agronomy10020214.

[36] H. Tsugawa *et al.*, 'A lipidome atlas in MS-DIAL 4', *Nat Biotechnol*, vol. 38, no. 10, pp. 1159–1163, Oct. 2020, doi: 10.1038/s41587-020-0531-2.

[37] H. Tsugawa *et al.*, 'MS-DIAL: Data-independent MS/MS deconvolution for comprehensive metabolome analysis', *Nat Methods*, vol. 12, no. 6, pp. 523–526, May 2015, doi: 10.1038/nmeth.3393.

[38] S. E. Stein and D. R. Scott, 'Optimization and testing of mass spectral library search algorithms for compound identification', *J Am Soc Mass Spectrom*, vol. 5, no. 9, pp. 859–866, 1994, doi: https://doi.org/10.1016/1044-0305(94)87009-8.

[39] Y. Li and O. Fiehn, 'Flash entropy search to query all mass spectral libraries in real time', *Nat Methods*, vol. 20, no. 10, pp. 1475–1478, Oct. 2023, doi: 10.1038/s41592-023-02012-9.